\documentclass[12pt]{article}
\pdfoutput=1
\usepackage{jheppub}
\usepackage{amssymb,amsmath,amstext,amsfonts}
\usepackage{bbold,ulem,bm}
\usepackage{graphics}
\usepackage{mathtools}
\usepackage{hyperref}
\usepackage{color}
\usepackage{graphicx}
\usepackage{subcaption}
\usepackage{mwe}
\usepackage{mathrsfs}
\usepackage{multirow}

\newcommand{\beq}{\begin{equation}}
\newcommand{\eeq}{\end{equation}}
\newcommand{\bea}{\begin{eqnarray}}
\newcommand{\eea}{\end{eqnarray}}
\newcommand{\be}{\begin{equation}}
\newcommand{\ee}{\end{equation}}
\newcommand{\bal}{\begin{aligned}}
\newcommand{\eal}{\end{aligned}}

\def\d{\mathrm{d}}

\begin{document}

\title{Gauged Galileons} 

\author[a]{Sebastian Garcia-Saenz,}
\affiliation[a]{Sorbonne Universit\'e, UPMC Paris 6 and CNRS, UMR 7095,
Institut d'Astrophysique de Paris, GReCO, 98bis boulevard Arago, 75014 Paris, France}
\author[b]{Jonghee Kang,}
\affiliation[b]{Center for Theoretical Physics and Department of Physics,
  Columbia University, \\ New York, NY 10027, USA}
\author[c]{Riccardo Penco}
\affiliation[c]{Department of Physics, Carnegie Mellon University, 5000 Forbes Ave, \\ Pittsburgh, PA 15217, USA}

%\emailAdd{sebastian.garcia-saenz@iap.fr}

\abstract{We discuss the gauging of non-linearly realized symmetries as a method to systematically construct spontaneously broken gauge theories. We focus in particular on galileon fields and, using a coset construction, we show how to recover massive gravity by gauging the galileon symmetry. We then extend our procedure to the special galileon, and obtain a theory that couples a massive spin-2 field with a traceless symmetric field, and is free of pathologies at quadratic order around flat space.}

\maketitle

%%%%%%%%%%%%%%%%%%%%%%%%%%%%%%%%%%%%%%%%%%%
%%%%%%%%%%%%%%%%%%%%%%%%%%%%%%%%%%%%%%%%%%%

\section{Introduction} \label{sec:intro}

The Goldstone boson equivalence theorem \cite{Cornwall:1974km,Lee:1977eg,Chanowitz:1985hj} states that the dynamics of massive gauge bosons is greatly simplified at energies far above the bosons' mass---but below the symmetry breaking scale---when a description based purely in terms of the interactions of the Goldstone scalars becomes applicable. An analogous story takes place in the context of theories of massive gravity. Even though the equivalent of the Higgs mechanism for spin-2 particles is unknown,\footnote{The concept of Higgs-type mechanism is used here in a loose sense, since it is likely that massive gravity does not possess any simple analogue of the traditional Higgs mechanism (see \textit{e.g.}~\cite{Bonifacio:2019mgk}). This of course does not invalidate the formalism of symmetry breaking and non-linear realizations as a tool to investigate theories of massive particles.} describing the spontaneously broken phase can still be achieved by means of the St\"uckelberg formulation. For a massive graviton, this amounts to the introduction of additional vector and scalar fields responsible for restoring diffeomorphism invariance, which is done while remaining agnostic about the physics behind the symmetry breaking. In this setting, at energies much higher than the graviton mass, a picture is recovered in which the relevant degrees of freedom correspond to massless spin-2, spin-1 and spin-0 particles, what is commonly referred to as the {\it decoupling limit} of massive gravity (see \cite{Hinterbichler:2011tt,deRham:2014zqa} for reviews).

Generic interactions for a massive graviton give rise to a pathological degree of freedom besides the expected five polarizations of a massive spin-2 particle (in four dimensions)---the infamous Boulware--Deser ghost \cite{Boulware:1973my}. In the language of effective field theory (EFT), this translates into a very low strong coupling scale that renders the model uninteresting from a phenomenological point of view. In the decoupling limit, the ghost manifests itself in the fact that the spin-0 equation of motion is of fourth order, producing an additional propagating mode that is unstable as a consequence of the Ostrogradsky theorem. As is well known, however, this issue can be solved by a judicious tuning of the graviton interactions, in what is known as the de Rham--Gabadadze--Tolley (dRGT) theory of massive gravity \cite{deRham:2010ik,deRham:2010kj,Hassan:2011hr,Hassan:2011tf,Hassan:2011ea}. The decoupling limit of this ghost-free theory is particularly enlightening \cite{Ondo:2013wka,Gao:2014ula}, with the spin-0 sector being described by a {\it galileon} theory \cite{Nicolis:2008in},\footnote{The galileon arises as an effective degree of freedom in several other settings besides that of massive gravity; see \cite{Trodden:2011xh,deRham:2012az,Deffayet:2013lga} for reviews.} a property that is at the heart of some of the virtues of the model, such as the Vainshtein mechanism \cite{Vainshtein:1972sx}.\footnote{Not only its virtues but also some of its vices, such as the problem of superluminal fluctuations; see {\it e.g.}~\cite{Padilla:2010tj,deFromont:2013iwa,Garcia-Saenz:2013gya}.}

In this paper we address what may be thought of as the opposite story to the decoupling limit: starting from a galileon theory for a single scalar field $\phi$, is it possible to systematically derive massive gravity as an {\it infrared} completion? We will answer this question in the affirmative---with some caveats. Our approach is based on the gauging of the symmetries that define the galileon, namely Poincar\'e invariance, a shift symmetry $\phi\to\phi+c$, and a ``galileon shift'' symmetry $\phi\to\phi+b_{\mu}x^{\mu}$.\footnote{This approach differs from that of \cite{Zhou:2011ix,Goon:2012mu} which considered the gauging of some additional symmetries under which the galileons transform linearly. This is also the case of the ``covariant galileon'' \cite{Deffayet:2009wt}, which can be seen to arise from the gauging of the linearly realized Poincar\'e symmetry.} That this is a natural starting point can be motivated by noting that this procedure works correctly in the simple case of a massive spin-1 field. There the decoupling limit theory is given by a scalar field $\pi$ and a shift invariance, with an action that depends on Lorentz invariant combinations of $\partial_{\mu}\pi$ (at lowest order in derivatives). Conversely, regarding the field $\pi$ as a Goldstone boson that non-linearly realizes a shift symmetry, gauging this symmetry amounts to introducing a 1-form gauge field $A_\mu$ while the shift is promoted to a local $U(1)$. The gauge theory is then constructed out of the invariant combination $\partial_{\mu}\pi+A_{\mu}$ according to the derivative expansion, yielding the action of a massive spin-1 field with $\pi$ now playing the role of a St\"uckelberg field. Returning to the galileon, in addition to the constant shift we now also have the galileon shift as a non-linearly realized symmetry. Upon gauging, the latter gives rise to a Lorentz vector-valued 1-form $h^a_{\phantom{a}\mu}$, which can naturally be interpreted as a vielbein for a dynamical spin-2 field. Indeed, just like the gauging of spacetime translations produces a massless spin-2 degree of freedom---a procedure that may be used to derive general relativity \cite{Delacretaz:2014oxa}---gauging the galileon shift, which may be better thought of as an ``internal'' translation, gives rise to a spin-2 field that is now generically massive, not being constrained by general coordinate invariance. Together with the gauge field associated to shifts and the original Goldstone scalar, we are thus left with all the necessary ingredients to construct theories of massive gravity in the St\"uckelberg formulation.

Now to the caveats. First, the galileon symmetry has a crucial difference relative to the usual constant shift; namely, it is a {\it spacetime} symmetry, in the sense that it does not commute with Poincar\'e transformations. Because of this, and the fact that gauge symmetries must form a subgroup, we will be forced to also gauge (at least part of) the unbroken Poincar\'e group, yielding in principle an {\it additional} local translation. In order to describe a massive spin-2 degree of freedom, we will have to explicitly break this symmetry by fixing the associated vielbein as a non-dynamical background field. Although this is a welcome extra ingredient that will allow us to formulate massive gravity in the general case of an arbitrary reference metric, it has the disadvantage that some of the symmetries must be broken by hand. Second, even though our formalism produces the correct degrees of freedom and symmetries, the interactions that we can construct are not limited to be the ghost-free ones of dRGT massive gravity. This is however an expected outcome given that the structure of the dRGT action is a result of a tuning of operators' coefficients in the derivative expansion, barring the existence of some hidden symmetry.

Motivated by the search of additional symmetries that could provide a rationale for the particular structure of dRGT theory, we also consider the gauging of the {\it special galileon} \cite{Hinterbichler:2015pqa}. The special galileon theory is a one-parameter subset of the generic galileon that enjoys an extended shift symmetry that is responsible for an enhanced soft behavior of scattering amplitudes \cite{Cheung:2014dqa}, among other interesting properties \cite{Novotny:2016jkh,Bogers:2018zeg,Roest:2019oiw}. In our formalism, the fact that this symmetry is spontaneously broken will allow for the presence of an additional Goldstone mode in addition to the massive spin-2 degrees of freedom that we are interested in. Remarkably, by imposing the absence of this extra field in the action at zeroth order in derivatives we will prove that the resulting potential for the graviton is of the ghost-free type, and that it is precisely the one-parameter subset of the dRGT action that maps onto the special galileon in the decoupling limit. In addition, we show that the two-derivative kinetic terms can also be engineered so as to decouple the extra Goldstone when linearized about flat spacetime. However, the question of whether a decoupling at the fully nonlinear level can be achieved remains open at this stage.

We emphasize that our method is completely general and systematic, and thus opens the door to several generalizations, even beyond the context of galileons and massive gravity. It is based on the extension of the coset construction to accommodate spontaneously broken gauge symmetries, which we review in Sec.\ \ref{sec:coset}. The application of the formalism to the galileon symmetry as a way to investigate theories of massive gravity is presented in Sec.\ \ref{sec:galileon}, and in Sec.\ \ref{sec:special galileon} we consider the extension of the symmetries to the case of the special galileon. We conclude with a discussion of our results in Sec.\ \ref{sec:discussion}. Some technical calculations are given in Appendix \ref{sec:app ghost free pot}. For pedagogical reasons we also present in some detail in Appendix \ref{sec:app special gal} the coset construction of the special galileon algebra.

\bigskip

\noindent
{\it Conventions:} We use the mostly-plus metric signature and work in $D$ spacetime dimensions, although we will specify some results to $D=4$. Coordinates of the spacetime manifold are labeled by greek indices (except in the general discussion of Sec.\ \ref{sec:coset}) and coordinates of the flat tangent space are labeled by latin indices. Symmetrization and antisymmetrization of indices is defined with unit weight. We will use anti-hermitian symmetry generators in order to avoid factors of $i$ in the algebras and in several other expressions.

%%%%%%%%%%%%%%%%%%%%%%%%%%%%%%%%%%%%%%%%%%%
%%%%%%%%%%%%%%%%%%%%%%%%%%%%%%%%%%%%%%%%%%%

\section{Coset construction for gauge symmetries} \label{sec:coset}

The coset construction \cite{Coleman:1969sm,Callan:1969sn} is a general and systematic method to derive the low-energy effective action for a set of Goldstone bosons associated with any given symmetry breaking pattern. This construction can also be used to describe the couplings between Goldstones and any additional field. While the Goldstones themselves transform nonlinearly under the broken symmetries, the coset construction furnishes a set of covariant building blocks which transform under {\it all} symmetries ({\it i.e.}, even the broken ones) according to linear representations of the unbroken group. Thus, by combining these elements in a way that is manifestly invariant under the unbroken symmetries, one can obtain the most general action that is invariant under all the symmetries. Moreover, the coset building blocks are usually organized in increasing order in derivatives, allowing one to produce invariant actions that systematically implement the derivative expansion. We refer the reader to \cite{Weinberg:1996kr} for a textbook treatment. 

The generalization of the original technique to accommodate spacetime symmetries ({\it i.e.}~symmetries that do not commute with the Poincar\'e group\footnote{One may of course consider background spacetimes other than Minkowski, in which case the Poincar\'e group would be replaced by the relevant isometry group. For simplicity, in this work we restrict our attention to Poincar\'e invariant theories.}) was developed in \cite{Volkov:1973vd, ogievetsky:1974ab}; some recent works that provide good self-contained reviews are \cite{Goon:2012dy,Nicolis:2013sga}. Here we limit ourselves to outline only the main elements of the coset construction, in order to establish our notation, and to review the method for including gauge symmetries in this set-up, following~\cite{Delacretaz:2014oxa}.\footnote{For a different but equivalent approach to gauge symmetries within the coset construction, see~\cite{Ivanov:1976pg,Goon:2014ika,Goon:2014paa}.} 

Let us consider a theory that is invariant under a symmetry group $G$, but with a ground state that spontaneously breaks $G$ down to a subgroup $H$. For simplicity, we will assume that spacetime translations are unbroken,\footnote{It would actually be sufficient for the discussion that follows to consider a static, homogeneous ground state. This weaker requirement amounts to demanding that there exists a set of generators $P_a$ that commute with each other and, if the system is also isotropic, have the correct transformation properties under rotations. Notice that the $P_a$'s could be a linear combination of the original space-time translations and internal generators---as is for instance the case in many condensed matter systems~\cite{Nicolis:2015sra}.} so that the algebra of $H$ is spanned by the generators $P_a$ of translations and by a set of generators $T_A$ that include all the other unbroken symmetries, both internal and spacetime, while the remaining generators of the algebra of $G$, the spontaneously broken ones, are denoted by $Z_\alpha$. We introduce the coset representative $\Omega(x)$, which is a $G$-valued field with no components along the unbroken generators $T_A$,
\beq
\Omega(x)\equiv e^{x^aP_a}e^{\pi^{\alpha}(x)Z_{\alpha}}\,,
\eeq
where the $\pi^\alpha$'s are the Goldstone fields associated with the broken generators $Z_\alpha$'s. From this, one defines the Maurer--Cartan form $\Omega^{-1}\d\Omega$, which is an algebra-valued field and as such may be expanded as a linear combination of all the generators:
\beq
\Omega^{-1}\d\Omega=e^aP_a+\omega^{\alpha}Z_{\alpha}+\omega^AT_A\,.
\eeq
The 1-forms $e^a$ are interpreted as vielbeins due to the way they transform under Lorentz and general coordinate transformations; in particular, one can define a metric $g_{\mu\nu}=\eta_{ab}e^a_{\phantom{a}\mu}e^b_{\phantom{b}\nu}$ with the expected transformation properties. The 1-forms $\omega^\alpha$ transform covariantly under all the symmetries, as announced, and are the basic building blocks out of which invariant actions may be constructed, simply by contracting indices with $H$-invariant tensors. It is in fact often more convenient to work with the Goldstone covariant derivatives $\nabla_a\pi^\alpha$, defined by
\beq
\nabla_a\pi^\alpha\equiv (e^{-1})_a^{\phantom{a}\mu}(\omega^{\alpha})_{\mu}\,.
\eeq
Lastly, the 1-forms $\omega^A$ transform as connections and are necessary in order to couple the Goldstones to matter fields, or to write higher-order covariant derivatives of the Goldstone fields themselves.

As is well known, in the presence of symmetry breaking patterns that involve spacetime symmetries, the standard counting of gapless modes dictated by Goldstone's theorem does not apply---see {\it e.g.}~\cite{Low:2001bw}. Instead, one finds in general that some of the Goldstones are redundant, in the sense that they are not necessary to achieve a non-trivial nonlinear realization of the symmetries under consideration. This usually happens either because these modes are not independent, or because they are gapped and may be integrated out upon restricting oneself to energy scales below the gap.\footnote{See however~\cite{Rothstein:2017twg} for a interesting exception, namely the {\it dynamical} Higgs mechanism.}
These redundant Goldstones can usually be eliminated directly at the level of the coset construction, by imposing the so-called ``inverse Higgs constraints'' \cite{Ivanov:1975zq}: whenever the algebra contains a commutator of the form $\left[P_a,Z^{\beta}\right]=\lambda_{a\alpha}^{\phantom{a\alpha}\beta}Z^{\alpha}+({\rm unbroken})$, then the constraint
\beq \label{eq:general ihc}
\lambda_{a\alpha}^{\phantom{a\alpha}\beta}\nabla^a\pi^{\alpha}=0\,,
\eeq
may be imposed and solved for the Goldstone $\pi^\beta$, which is redundant. The solution can then be substituted back into the components of the Maurer--Cartan form without affecting their transformation properties.\footnote{We refer the reader to \cite{Nicolis:2013sga,Brauner:2014aha,Klein:2017npd} for more complete discussions related to redundant Goldstones and inverse Higgs constraints, as well as their physical interpretation.}

The discussion so far has been restricted to global symmetries, but the coset construction can be easily generalized to the case where some of the symmetries are gauged, be they broken or unbroken by the ground state \cite{Delacretaz:2014oxa}. Let $G_g\subseteq G$ be the gauged subgroup, and denote its generators by $V_I$. The inclusion of gauge fields $A^I(x)$ is done by modifying the Maurer--Cartan form as
\beq
\Theta\equiv \Omega^{-1}\left(\d+A^IV_I\right)\Omega\,.
\eeq
One can easily verify that $\Theta$ is invariant under local $G_g$ transformations, 
\beq
\Omega\to g(x)\Omega\,,\qquad A^IV_I\to g(x)A^IV_I g^{-1}(x)+g(x)\d g^{-1}(x)\,,
\eeq
with $g(x)\in G_g$.

As usual, the gauge invariance related to a spontaneously broken gauge symmetry can be used to completely eliminate the corresponding Goldstone field, by working in the so-called unitary gauge. To see this explicitly in our set-up, consider a subgroup $G_g'\subseteq G_g$ of gauged generators $V_I$ which are also broken, with associated Goldstone fields $\pi^I$. To leading order in the Goldstones we have
\beq
\omega^I=A^I+\d\pi^I+f^{I}_{\phantom{I}JK}A^J\pi^K+O(\pi^2)\,,
\eeq
with $f^{I}_{\phantom{I}JK}$ the structure constants of the algebra of $G_g'$. On the other hand, under an infinitesimal gauge transformation $g(x)=e^{\epsilon^I(x)V_I}\simeq 1+\epsilon^IV_I$, the gauge field changes as $A^I\to A^I-\d\epsilon^I-f^{I}_{\phantom{I}JK}A^J\epsilon^K$, so that the choice of gauge $\epsilon^I=\pi^I$ precisely eliminates the Goldstones $\pi^I$ from the Maurer--Cartan form. Although obviously more economic, the language of unitary gauge is not necessarily the most useful one, since depending on the context it may obscure the correct power counting of operators in the derivative expansion.\footnote{This is completely analogous to the treatment of massive field theories in the St\"uckelberg formulation, see {\it e.g.} \cite{deRham:2018qqo,Boulanger:2018dau}.} For this reason, it is in general good practice to carry out the coset construction keeping all the Goldstones and gauge fields, and only fix unitary gauge after building the action, if so desired.

The Maurer--Cartan 1-form provides all the necessary ingredients to build kinetic terms for the Goldstones. On the other hand, kinetic terms for the gauge fields are furnished by the components of
\beq\bal
\Theta_2&\equiv \Omega^{-1}\left(\d+A^IV_I\right)^2\Omega = \Omega^{-1}\left(2\d A^IV_I+A^I\wedge A^J[V_I,V_J]\right)\Omega\,,
\eal\eeq
which we will refer to as the Maurer--Cartan 2-form. Because the gauged generators form an algebra, with some structure constants $f^{I}_{\phantom{I}JK}$, we can write $\Theta_2$ as
\beq
\Theta_2=\Omega^{-1}\left(2\d A^I+f^{I}_{\phantom{I}JK}A^J\wedge A^K\right)V_I\Omega\,,
\eeq
and we recognize inside the parenthesis the usual gauge field strength, but which in the spontaneously broken case will in general receive corrections proportional to the Goldstones.

%%%%%%%%%%%%%%%%%%%%%%%%%%%%%%%%%%%%%%%%%%%
%%%%%%%%%%%%%%%%%%%%%%%%%%%%%%%%%%%%%%%%%%%

\section{Gauged galileons and massive gravity} \label{sec:galileon}

In this section we carry out the coset construction of the gauged galileon algebra and investigate its relation to theories of massive gravity. 

\subsection{Coset Construction}

In addition to the generators of translations $P_a$ and Lorentz transformations $J_{ab}$, the galileon algebra contains a constant shift generator $C$ and some internal translations---or``galileon shifts''---generated by $Q_a$. On a scalar field $\pi$, the latter correspond to symmetry transformations $\delta_C \pi=1$ and $\delta_{Q_a}\pi=x_a$. The non-vanishing commutators of the algebra are given by
\beq\bal \label{eq:gal algebra}
[P_a,Q_b]&=\eta_{ab}C\,,\\
[J_{ab},Q_c]&=\eta_{bc}Q_a-\eta_{ac}Q_b\,,\\
[J_{ab},P_c]&= \eta_{bc} P_a - \eta_{ac} P_b\,,\\
[J_{ab},J_{cd}]&=\eta_{bc} J_{ad} - \eta_{ac} J_{bd} - \eta_{bd} J_{ac} + \eta_{ad} J_{bc}\,.
\eal\eeq
with the last two making the usual Poincar\'e subalgebra. The Galileon theory is defined around a state that breaks the symmetry generators $C$ and $Q_a$ down to Poincar\'e \cite{Goon:2012dy}, so that the coset representative we consider is
\beq
\Omega=e^{x^aP_a}e^{\pi C}e^{\xi^bQ_b}\,,
\eeq
with Goldstones $\pi(x)$ and $\xi^a(x)$. Because we are interested in obtaining a spontaneously broken gauge theory, and because the gauged generators must form a subalgebra, we conclude that there are two possible ways to include gauge symmetries in this context: either by gauging $P_a$, $Q_a$ and $C$, or by gauging the whole algebra. However, the first case without gauging $J_{ab}$ would not give all the necessary ingredients that we expect in a sensible theory of gravity. Indeed, in the case of pure Poincar\'e it is necessary to gauge both $P_a$ and $J_{ab}$ in order to derive General Relativity \cite{Delacretaz:2014oxa}. We therefore focus on the case where all the symmetries are gauged, and introduce 1-form gauge fields $\tilde{e}^a$, ${\omega}^{ab}$, $\tilde{A}$ and ${h}^a$ respectively for translations, Lorentz transformations, shifts and galileon shifts. The Maurer--Cartan form is then
\beq\bal
\Theta&\equiv\Omega^{-1}\left(\d+\tilde{e}^aP_a+\frac{1}{2}\,{\omega}^{ab}J_{ab}+{h}^aQ_a+\tilde{A}C\right)\Omega\\
&\equiv e^aP_a+\frac{1}{2}\,\omega^{ab}J_{ab}+\omega_Q^aQ_a+\omega_C C\,,
\eal\eeq
where the covariant 1-forms are given by
\beq\bal \label{eq:gal 1-forms}
e^a&=\delta^a+\tilde{e}^a+{\omega}^{ab}x_b\,,\\
\omega_Q^a&=\d\xi^a+\omega^{ab}\xi_b+h^a\equiv e^b_{\phantom{b}\mu}\nabla_b\xi^a\,\d x^{\mu}\,,\\
\omega_C&=\d\pi+A+e^a\xi_a\equiv e^b_{\phantom{b}\mu}\nabla_b\pi\,\d x^{\mu}\,,\\
\eal\eeq
with $\delta^a\equiv \delta^a_{\mu}\d x^{\mu}$ and $A\equiv \tilde{A}-{h}^ax_a$. By performing simple field redefinitions, we can now consider $e^a$, $\omega^{ab}$, $h^a$ and $A$ as the relevant variables for our gauge fields, and recall that $e^a$ and $\omega^{ab}$ define, respectively, the vielbein and spin connection of the spacetime manifold.

It is convenient at this stage to eliminate the redundant Goldstones by identifying the inverse Higgs constraints. The algebra commutator $[P_a,Q_b]=\eta_{ab}C$ allows us to set (\textit{cf.}~\eqref{eq:general ihc})
\beq
\nabla_a\pi=0\qquad \Rightarrow\qquad \xi_a=-\partial_a\pi-A_a\,,
\eeq
where as usual we employed the (inverse) vielbein to trade Lorentz for spacetime indices, {\it e.g.}~$A_a\equiv(e^{-1})_a^{\phantom{a}\mu}A_{\mu}$. Notice that, after solving the constraint, the choice of unitary gauge will allow us to eliminate both $\pi$ and $A$ from the action.

The last ingredients we need are the gauge field strengths, which are necessary to build kinetic terms and derivative interactions for the gauge fields. The Maurer--Cartan 2-form is
\beq\bal
\Theta_2&=\Omega^{-1}\left(\d+\tilde{e}^aP_a+\frac{1}{2}\,{\omega}^{ab}J_{ab}+{h}^aQ_a+\tilde{A}C\right)^2\Omega\\
&=T^aP_a+\frac{1}{2}\,R^{ab}J_{ab}+T_Q^aQ_a+T_CC\,,
\eal\eeq
and we find the components
\beq\bal \label{eq:gal 2-forms}
T^a&=\d e^a-e^b\wedge\omega^a_{\phantom{a}b}\,,\\
T_Q^a&=\d\omega_Q^a-\omega_Q^b\wedge\omega^a_{\phantom{a}b}\,,\\
T_C&=\d\omega_C+e_a\wedge\omega_Q^a\,,\\
R^{ab}&=\d\omega^{ab}+\omega^{ac}\wedge\omega_c^{\phantom{c}b}\,.
\eal\eeq
The last term is the familiar Riemann 2-form for the spin connection. The 2-form associated with $C$ does not carry any new information, since it reduces to $T_C=e_a\wedge\omega_Q^a$ upon solving the inverse Higgs constraint, which we already knew to be a covariant object. Lastly, the 2-forms associated with $P_a$ and $Q_a$ correspond to torsion fields for $e^a$ and $\omega_Q^a$.

\subsection{Background Vielbein and Torsion-free Condition}

The 1-forms $e^a$ and $\omega_Q^a$ in \eqref{eq:gal 1-forms} and the 2-forms $T^a$, $T_Q^a$ and $R^{ab}$ in \eqref{eq:gal 2-forms} give all the building blocks required to construct invariant actions at lowest order in derivatives. At this stage we apply some physical input in view of the type of theories we seek, {\it i.e.}~theories with a massive spin-2 field. Although both $e^a$ and $h^a$---or equivalently $\omega_Q^a$---contain a rank-2 symmetric tensor (upon pulling back to the spacetime manifold), physically we cannot regard both as dynamical because with a single Riemann 2-form we can only write down one Einstein--Hilbert kinetic term. Recall that in general relativity and massive gravity this is done by expressing the spin connection $\omega^{ab}$ in terms of the vielbein after imposing that the torsion vanish. Here we have two torsions and one spin connection, so the best we can do is to set a particular combination of $T^a$ and $T_Q^a$ to zero, but not both. To identify this combination, the first step is to define the vielbein that is going to describe the spin-2 degree of freedom in our theory. A natural choice is
\beq
q^a\equiv e^a+\omega_Q^a\,,
\eeq
and we will interpret $e^a$ to be a {\it non-dynamical} background vielbein, and $\omega_Q^a$ as the fluctuation field about this background. That this choice makes sense can be seen by noting that, in unitary gauge where the Goldstones are set to zero, the ``vacuum'' configuration in which the gauge fields vanish corresponds to
\beq
e^a=\delta^a\,,\qquad \omega_Q^a=0\,,
\eeq
so that the above interpretation is consistent. Of course, since $e^a$ was originally the gauge field of local translations, by treating it as an externally prescribed field we are giving up diffeomorphism invariance, which is again consistent with our goal of constructing models of massive gravity.

Another choice we will make on physical grounds is to eliminate the spin connection via some torsion-free condition. Having in mind that the kinetic interactions of $q^a$ should be given by the Einstein--Hilbert term,\footnote{Ghost-free massive spin-2 kinetic interactions that are not of the Einstein--Hilbert type have been ruled out in \cite{deRham:2013tfa,deRham:2015rxa}. This result is reminiscent of what happens for massive spin-1 fields, where the only ghost-free kinetic term is the gauge-invariant one. \label{footnote kinetic terms}} we impose the condition
\beq
T^a+T_Q^a=\d q^a-q^b\wedge\omega^a_{\phantom{a}b}=0\,,
\eeq
which has the standard solution for the spin connection,
\beq \label{eq:spin connection sol}
\omega^{ab}_{\phantom{ab}\mu}=(q^{-1})^{\rho[a}\partial_{\mu}q_{\rho}^{\phantom{\rho}b]}-(q^{-1})^{\lambda[a}\partial_{\rho}q_{\mu}^{\phantom{\rho}b]}-(q^{-1})^{\rho[a}(q^{-1})^{b]\sigma}q_{\mu c}\partial_{\rho}q_{\sigma}^{\phantom{\sigma}c}\,.
\eeq

In summary, we have eliminated $e^a$ and $\omega^{ab}$ as physical variables by choosing to give up diffeomorphism invariance and by imposing the condition of vanishing torsion. In addition, and as mentioned before, the Goldstone $\pi$ and the gauge field $A$---which may be better thought of as a Goldstone or St\"uckelberg field after the inverse Higgs constraint is imposed---are pure gauge degrees of freedom and may be fully eliminated by choosing to work in unitary gauge. This leaves us with $h^a$ as the only propagating field, and with local Lorentz transformations as the only relevant symmetry. These are precisely the defining properties of a theory of massive gravity.

\subsection{Effective Action}

The most general action can now be straightforwardly constructed through a standard derivative expansion. However, it is well known that generic interactions for a massive spin-2 particle lead to an additional ghostly degree of freedom \cite{Boulware:1973my}, which can only be remedied by a special tuning of operator coefficients.\footnote{Of course, in an effective field theory the presence of a ghost simply signals an incorrect identification of the UV cutoff. The requirement of ``having no ghosts'' is thus a phenomenological one, motivated by the need of enlarging the range of applicability of the theory.} This leads to the dRGT theory of massive gravity \cite{deRham:2010ik,deRham:2010kj}:\footnote{The prefactor has been chosen so that the Einstein--Hilbert term yields the standard normalization when switching to the metric formulation: $S_{\rm EH}=\frac{M_P^{D-2}}{2}\int d^Dx\sqrt{-g}\,R(g)$.}
\beq\bal \label{eq:drgt action}
S_{\mathrm{dRGT}}&=\frac{M_P^{D-2}}{6(D-2)!}\Bigg[\int \epsilon_{a_1a_2\cdots a_D}R^{a_1a_2}(q)\wedge q^{a_3}\wedge\cdots\wedge q^{a_D}\\
&\quad +m^2\int\sum_{n=0}^{D-1}\frac{c_n}{n!(D-n)!}\,{\epsilon}_{a_1a_2\cdots a_D}e^{a_1}\wedge\cdots\wedge e^{a_n}\wedge q^{a_{n+1}}\wedge\cdots\wedge q^{a_D}\Bigg]\,,
\eal\eeq
with $M_P$ the Planck mass and $m$ the graviton mass. The Riemann form is now a function of $q^a$, so that the first term is the standard Einstein--Hilbert action in second-order form. Among the $D$ dimensionless coefficients $c_n$, one combination of them is fixed after $m$ is chosen as the physical graviton mass, and another combination can be set to zero by assuming the absence of a tadpole term. This leaves $D-2$ independent parameters in addition to $M_P$ and $m$.

It is worth repeating that the action \eqref{eq:drgt action} is a result of a tuning of EFT coefficients in our formalism and does not follow from a derivative expansion. We have shown that the coset construction of the gauged galileon group provides all the necessary ingredients to build theories of massive gravity, but ultimately no additional symmetries are present that can be used to restrict the set of allowed interactions. This outcome was expected, as it is known that the form of the dRGT action is not protected against under quantum corrections \cite{deRham:2013qqa}, suggesting that it is not protected by any symmetry. An analogous conclusion was reached in~\cite{Goon:2014paa}, where a similar line of reasoning---although in a different set-up---was used to construct massive gravity as a spontaneously broken gauge theory. 

%%%%%%%%%%%%%%%%%%%%%%%%%%%%%%%%%%%%%%%%%%%
%%%%%%%%%%%%%%%%%%%%%%%%%%%%%%%%%%%%%%%%%%%

\section{Gauged special galileons} \label{sec:special galileon}

The special galileon algebra extends the galileon algebra \eqref{eq:gal algebra} with a generator $S_{ab}$ that is symmetric and traceless in its Lorentz indices \cite{Hinterbichler:2015pqa}. It is realized on a scalar field $\phi$ as an extended shift symmetry,
\beq \label{special Galileon symmetry}
\delta_{S_{ab}}\pi=x_ax_b-\frac{1}{D}\,\eta_{ab}x^2-\alpha^2\left[\partial_a\pi\partial_b\pi-\frac{1}{D}\,\eta_{ab}(\partial\pi)^2\right]\,,
\eeq
where $\alpha$ is an arbitrary constant. In what follows, we will set $\alpha=1$ without loss of generality by a rescaling of generators. We will now repeat the analysis in the previous section for a Galileon field endowed with the additional symmetry \eqref{special Galileon symmetry}.

\subsection{Coset Construction}

In addition to \eqref{eq:gal algebra}, the special galileon algebra contains the following non-trivial commutators:
\beq\bal
\left[P_a,S_{bc}\right]&=\eta_{ab}Q_c+\eta_{ac}Q_b-\frac{2}{D}\,\eta_{bc}Q_a\,,\\
[Q_a,S_{bc}]&=\eta_{ab}P_c+\eta_{ac}P_b-\frac{2}{D}\,\eta_{bc}P_a\,,\\
[S_{ab},S_{cd}]&=\eta_{bc} J_{ad}+\eta_{ac} J_{bd} + \eta_{bd} J_{ac} + \eta_{ad} J_{bc}\,,\\
[J_{ab},S_{cd}]&=\eta_{bc}S_{ad}-\eta_{ac}S_{bd}+\eta_{bd}S_{ac}-\eta_{ad}S_{bc}\,.
\eal\eeq
The special galileon theory is defined by a state that breaks $C$, $Q_a$ and $S_{ab}$ down to the Poincar\'e subgroup.

Our goal is to work out the coset construction for the gauged special galileon. Goldstone bosons are introduced via the coset parametrization
\beq
\Omega=e^{x^aP_a}e^{\pi C}e^{\xi^bQ_b}e^{\frac{1}{2}\sigma^{cd}S_{cd}}\,,
\eeq
where $\sigma^{ab}$ is symmetric and traceless. The algebra contains three subalgebras: Poincar\'e, the subalgebra formed by $P_a$, $Q_a$ and $C$, and the galileon algebra \eqref{eq:gal algebra}. When it comes to choose which symmetries to gauge, we ignore as before the options of gauging only Poincar\'e (because we are interested in broken gauge symmetries), or only $P_a$, $Q_a$ and $C$ (because of the absence of a spin connection). This leaves us with the possibilities of gauging either the whole group or the galileon subgroup. We make the simplest choice of gauging only the galileon subgroup, which as we will see will allow us to recover a subset of the dRGT potentials. Nevertheless, the option of gauging the extended symmetry $S_{ab}$ is interesting and we will comment on it in Sec.\ \ref{sec:discussion}. 

The Maurer--Cartan form is therefore given by
\beq\bal
\Theta&=\Omega^{-1}\left(\d+\tilde{e}^aP_a+\frac{1}{2}\,{\omega}^{ab}J_{ab}+{h}^aQ_a+\tilde{A}C\right)\Omega\\
&=E^aP_a+\frac{1}{2}\,\Omega^{ab}J_{ab}+\Omega_Q^aQ_a+\Omega_CC+\frac{1}{2}\,\Omega_S^{ab}S_{ab}\,,
\eal\eeq
where $E^a$ and $\Omega^{ab}$ (not to be confused with the coset representative) are the vielbein and spin connection, and $\Omega^a_Q$, $\Omega_C$ and $\Omega_S^{ab}$ are the covariant 1-forms associated to the broken generators. In terms of the redefined gauge fields $e^a\equiv \delta^a+\tilde{e}^a+{\omega}^{ab}x_b$ and $A\equiv\tilde{A}-{h}^ax_a$ introduced in the previous section, we find\footnote{The calculation of $\Omega^{ab}$ and $\Omega_S^{ab}$ is non-trivial; the interested reader can find more details in Appendix~\ref{sec:app special gal}.}
\beq \label{MC 1 forms special galileon}\bal
E^a&=(\cosh\sigma)^{a}_{\phantom{a}b}e^b+(\sinh\sigma)^{a}_{\phantom{a}b}\omega_Q^b\,,\\
\Omega^{ab}&=\omega^{ab}+(\Lambda^{-1})^{ab}_{cd}\left(\cosh\Lambda-\mathbf{1}\right)^{cd}_{ef}\left(\d\sigma^{ef}+2\omega^{g(e}\sigma^{f)}_{\phantom{f)}g}\right) \,,\\
\Omega_Q^a&=(\sinh\sigma)^{a}_{\phantom{a}b}e^b+(\cosh\sigma)^{a}_{\phantom{a}b}\omega_Q^b\equiv E^b_{\phantom{b}\mu}\nabla_b\xi^a\,\d x^{\mu}\,,\\
\Omega_C&=\d\pi+A+e^a\xi_a\equiv E^a_{\phantom{a}\mu}\nabla_a\pi\,\d x^{\mu}\,,\\
\Omega_S^{ab}&=(\Lambda^{-1})^{ab}_{cd}\left(\sinh\Lambda\right)^{cd}_{ef}\left(\d\sigma^{ef}+2\omega^{g(e}\sigma^{f)}_{\phantom{f)}g}\right) \equiv E^c_{\phantom{c}\mu}\nabla_c\sigma^{ab}\,\d x^{\mu}\,.
\eal\eeq
where we defined $\Lambda^{ab}_{cd}\equiv \delta^a_c\sigma^b_{\phantom{b}d}-\delta^b_d\sigma^a_{\phantom{a}c}$. Notice that $E^a, \Omega^{a b}$ and $\Omega^a_Q$ respectively reduce to the quantities $e^a, \omega^{a b}$ and $\omega^a_Q$ introduced in the previous section when $\sigma^{ab} = 0$.

The inverse Higgs constraint related to the galileon algebra commutator $[P_a,Q_b]=\eta_{ab}C$ is exactly the same as in Sec.\ \ref{sec:galileon}, {\it i.e.}
\beq
\nabla_a\pi=0\qquad \Rightarrow\qquad \xi_a=-\partial_a\pi-A_a\,.
\eeq
In the special galileon case there is also another constraint related to the commutator $[P_a,S_{bc}]=\eta_{ab}Q_c+\eta_{ac}Q_b-\frac{2}{D}\,Q_a\eta_{bc}$. This implies that we can set to zero the traceless symmetric part of the covariant derivative of $\xi^a$,
\beq \label{eq:second ihc}
\nabla_a\xi_b+\nabla_b\xi_a-\frac{2}{D}\,\eta_{ab}[\nabla\xi]=0\,,
\eeq
where $[\ldots]$ stands for the trace. This equation can in principle be solved to express $\sigma_{ab}$ in terms of derivatives of the other Goldstones and the gauge fields.

Let us now define the rank-2 tensor $K_a^{\phantom{a}b}\equiv (e^{-1})_a^{\phantom{a}\mu}(\omega_Q^b)_\mu$. This is essentially the covariant derivative of the Goldstones $\xi^a$ in the absence of the field $\sigma^{a b}$---see Eq.\ \eqref{eq:gal 1-forms}.
Switching for a moment to a matrix notation with contractions of Lorentz indices denoted by a dot, we can write $E=e\cdot(\cosh\sigma+K\cdot\sinh\sigma)$ and $\omega_Q=e\cdot(\sinh\sigma+K\cdot\cosh\sigma)$, and therefore
\beq
\nabla\xi=(\cosh\sigma+K\cdot\sinh\sigma)^{-1}\cdot (\sinh\sigma+K\cdot\cosh\sigma)\,.
\eeq
The inverse Higgs constraint \eqref{eq:second ihc} now implies a relation between $\sigma$ and $K$, which means that a solution exists only if $[\sigma,K]=0$. This observation allows us to rewrite $\nabla\xi$ as
\beq \label{eq:nabla xi matrix}
\nabla\xi=\tanh(\sigma+\tanh^{-1}K)\,.
\eeq
At this point we will make the additional assumption that $\nabla\xi$ is a symmetric matrix, {\it i.e.}~$\nabla_a\xi_b=\nabla_b\xi_a$. We will expand on this assumption below, where we will show that it is related to the so-called ``symmetric vielbein condition''~\cite{Deser:1974cy} commonly used in the context of massive gravity.

Using such symmetry condition in \eqref{eq:second ihc} we find that $\nabla\xi=\frac{\mathbf{1}}{D}\,[\nabla\xi]$, which we substitute in \eqref{eq:nabla xi matrix} to obtain (reinstating indices)
\beq \label{eq:second ihc solution}
\sigma^{ab}=-(\tanh^{-1}K)^{ab}+\frac{1}{D}\,\eta^{ab}[\tanh^{-1}K]\,.
\eeq
This is the desired solution of the second inverse Higgs constraint, which defines $\sigma^{ab}$ as a function of the Goldstone $\pi$ and the gauge fields.

\subsection{Effective Action: potential terms}

It is worth pausing at this stage to consider the ingredients we have so far, and the type of invariant operators we can build out of them. The relevant fields are again $e^a$ and $\omega_Q^a$ (which reduces to $h^a$ in unitary gauge), the spin connection (which due to a torsion-free condition will ultimately not be independent), and now possibly also $\sigma^{ab}$ (depending on whether or not we choose to apply the second inverse Higgs constraint to express $\sigma^{ab}$ in terms of $e^a$ and $\omega_Q^a$ as in Eq.\ \eqref{eq:second ihc solution}). Keeping in mind our goal of constructing theories of massive gravity, we will remove the diffeomorphism symmetry associated to the gauging of translations by setting $e^a$ to be a non-dynamical background field, and interpret $\omega_Q^a$ as the spin-2 fluctuation about this background---precisely as we did in the standard galileon case. The difference is that now $e^a$ and $\omega_Q^a$ do not transform covariantly under the special galileon group, but instead the correct building blocks are now $E^a$ and $\Omega_Q^a$ defined in Eq.\ \eqref{MC 1 forms special galileon}. In particular, the dRGT type potentials we are interested in, namely the terms of the form
\beq \label{eq:generic drgt pot}
S_{\rm pot}=\int\sum_{n=0}^D\frac{b_n}{n!(D-n)!}\,{\epsilon}_{a_1a_2\cdots a_D}\Omega_Q^{a_1}\wedge\cdots\wedge\Omega_Q^{a_n}\wedge E^{a_{n+1}}\wedge\cdots\wedge E^{a_D}\,,
\eeq
will generically involve interactions between $\omega_Q^a$ and the Goldstone $\sigma^{ab}$, as well as self-interactions for the latter. If we choose to replace $\sigma^{ab}$ in terms of $\omega_Q^a$ by solving the inverse Higgs constraint, Eq.\ \eqref{eq:second ihc solution}, the resulting potential will clearly not be of the ghost-free type. If instead we decide not to apply the constraint and leave $\sigma^{ab}$ as an independent degree of freedom, which will generically be gapped, the action will now yield the correct self-interactions for $\omega_Q^a$, but also some extra operators involving the Goldstone $\sigma^{ab}$, and it is far from clear whether they will be free of pathologies.

Remarkably, however, we can bypass these issues---at least as far as the potential is concerned---by observing that it is possible to choose the coefficients $b_n$ in \eqref{eq:generic drgt pot} in such a way that $\sigma^{ab}$ drops out completely from the expression in \eqref{eq:generic drgt pot}. In Appendix \ref{sec:app ghost free pot}, we show in fact  that for the specific choice
\beq
b_n=\begin{cases}
\beta_1 & \mbox{if $n$ is even}\,,\\
\beta_2 & \mbox{if $n$ is odd}\,,
\end{cases}
\eeq
with $\beta_1$ and $\beta_2$ arbitrary constants, the action in \eqref{eq:generic drgt pot} reduces to
\beq\bal \label{eq:special drgt pot}
S_{\rm pot}&=\int \,{\epsilon}_{a_1a_2\cdots a_D} \Bigg[  \sum_{n\,{\mathrm{even}}}\frac{\beta_1}{n! (D-n)!}\,  e^{a_1}\wedge\cdots\wedge e^{a_n}\wedge \omega_Q^{a_{n+1}}\wedge\cdots\wedge \omega_Q^{a_D} \\
&\quad + \sum_{n\,{\mathrm{odd}}} \frac{\beta_2}{n! (D-n)!}\, e^{a_1}\wedge\cdots\wedge e^{a_n}\wedge \omega_Q^{a_{n+1}}\wedge\cdots\wedge \omega_Q^{a_D} \Bigg]\,.
\eal\eeq
This action has the desired ghost-free property, being a specific member of the dRGT family of potentials. However, the $n=1$ term in the above sum gives rise to a tadpole, and therefore we will set $\beta_2=0$ in what follows. The parameter $\beta_1$ instead is related to the graviton mass as explained in Sec.\ \ref{sec:galileon}. Thus, in this theory, the graviton mass is the unique parameter controlling the non-derivative spin-2 interactions. In relation to our original motivation of constructing IR completions of the special galileon, one can check that dRGT massive gravity with the ``special'' potential \eqref{eq:special drgt pot} indeed yields the special galileon theory in its decoupling limit.\footnote{Albeit with some additional interactions with helicity-2 modes.}

Incidentally, although we have been able to engineer an action in which the Goldstone $\sigma^{ab}$ plays no role, it is worth going back to the assumption we made when solving the second inverse Higgs constraint, namely the symmetry condition $\nabla_a\xi_b=\nabla_b\xi_a$. Remembering that $\nabla_b\xi^a=(E^{-1})_b^{\phantom{b}\mu}(\Omega_Q^a)_{\mu}$, we observe that the symmetry of $\nabla_a\xi_b$ is equivalent to the so-called symmetric vielbein condition that is typically used in the context of ghost-free massive gravity, and which allows to connect the vielbein and metric formulations. It has been shown that, for the dRGT type potentials \eqref{eq:generic drgt pot}, this condition is not an extra assumption but actually follows from the equations of motion \cite{Hinterbichler:2012cn}.\footnote{It was later pointed out in \cite{Deffayet:2012zc} that this is not strictly true in general, as there exist non-perturbative field configurations for which the argument may fail depending on the parameters.} Precisely the same reasoning can be used in our case if we decide to focus on ghost-free potentials.

\subsection{Effective Action: kinetic terms}

Lastly, we turn our attention to the kinetic terms which are generated by the coset construction. The Maurer--Cartan 2-form is given by
\beq\bal
\Theta_2&=\Omega^{-1}\left(\d+\tilde{e}^aP_a+\frac{1}{2}\,{\omega}^{ab}J_{ab}+{h}^aQ_a+\tilde{A}C \right)^2\Omega\\
&\equiv{\cal T}^aP_a+\frac{1}{2}\,{\cal R}^{ab}J_{ab}+{\cal T}_Q^aQ_a+{\cal T}_CC + \frac{1}{2}\,{\cal T}_S^{ab} S_{ab}\,,
\eal\eeq
and the various components are
\beq \label{MC 2 forms special galileon}\bal
{\cal T}^a&=\d E^a-E^b \wedge\Omega^a_{\phantom{a} b} + \Omega_Q^b \wedge{\Omega_S}^a_{\phantom{a} b}  \,,\\
{\cal R}^{ab}&=\d\Omega^{ab}+\Omega^a_{\phantom{a} c}\wedge\Omega^{cb}+{\Omega_S}^a_{\phantom{a} c}\wedge{\Omega_{S}}^{cb}\,,\\
{\cal T}_Q^a&=\d\Omega_Q^a-\Omega_Q^b \wedge\Omega^a_{\phantom{a} b} + E^b \wedge{\Omega_S}^a_{\phantom{a} b} \,,\\
{\cal T}_C&=\d\Omega_C+E_a \wedge\Omega_Q^a\,,\\
{\cal T}_S^{ab} &= \d\Omega_S^{ab}+\Omega^a_{\phantom{a} c}\wedge{\Omega_S}^{cb}+\Omega^b_{\phantom{a} c}\wedge{\Omega_S}^{ca}\, .
\eal\eeq
The tensors ${\cal T}^a$ and ${\cal R}^{ab}$ generalize the spacetime torsion and Riemann 2-forms to include additional contributions from the Goldstone $\sigma^{ab}$. The ``$Q$-torsion'' ${\cal T}^a_Q$ likewise involves extra terms proportional to $\sigma^{ab}$, while ${\cal T}_C$ is again redundant upon solving the first inverse Higgs constraint.

At this stage we are faced again with the issue of constructing a non-pathological action for the spin-2 field $\omega_Q^a$ (or, in unitary gauge, $h^a$). When considering non-derivative operators in the previous subsection, we had a set of coefficients that could be tuned so as to yield a result that was independent of $\sigma^{ab}$, which was desirable in view of the complications related to this additional field. Now, the potential in Eq.\ \eqref{eq:special drgt pot} implies the absence of any mass terms for $\sigma^{ab}$ (which, being a redundant Goldstone, should generically be present). Therefore, the application of the inverse Higgs constraint, albeit still valid from a symmetry viewpoint, is no longer equivalent to integrating out $\sigma^{ab}$ at the level of the action. For this reason, we proceed by considering $\sigma^{ab}$ as an independent degree of freedom, in addition to $\omega^a_Q$.

Focusing on the kinetic self-interactions of the spin-2 field $\omega_Q^a$, a natural guess for the kinetic term would be
\beq \label{eq:special gal eh term}
S_{\rm EH} = \frac{M_P^{D-2}}{6(D-2)!}\int {\epsilon}_{a_1\,a_2 \cdots a_D} {\cal R}^{a_1 a_2} \wedge q^{a_3}  \wedge \cdots \wedge q^{a_D}\, ,
\eeq
where $q^a\equiv E^a+\Omega_Q^a$. We would like this to reduce to the Einstein--Hilbert action for the vielbein $e^a+\omega_Q^a$ when $\sigma^{ab}$ is turned off, which implies that the correct torsion-free condition we should impose is
\beq
{\cal T}^a+{\cal T}^a_Q=\d q^a - q^b \wedge \Omega^a_{\phantom{a} b} + q^b \wedge {\Omega_S}^a_{\phantom{a} b} \equiv 0 \, .
\eeq
This can be solved for the spin connection as
\beq \label{eq:special spin connection sol}
\Omega^{ab}_{\phantom{ab}\mu}=\bar{\Omega}^{ab}_{\phantom{ab}\mu}(q) +  2  q_{\mu c} (q^{-1}) ^{\rho [a} \Omega_{S\phantom{ b] c}\rho}^{\phantom{S} b] c} \, ,
\eeq
where $\bar{\Omega}^{ab}_{\phantom{ab}\mu}(q)$ is the usual solution for the spin connection as a function of $q^a$ ({\it cf.}\ \eqref{eq:spin connection sol}).

While $S_{\rm EH}$ indeed generates the correct kinetic term for $\omega_Q^a$, it also induces self and mutual derivative interactions for the Goldstone $\sigma^{ab}$ (notice from Eq.\ \eqref{MC 1 forms special galileon} that $\Omega_S^{ab}\simeq\d\sigma^{ab}$ at lowest order in $\sigma^{ab}$). In fact, in the absence of an analogue of the Einstein--Hilbert term for $\sigma^{ab}$ (notice that the ``curvature'' ${\cal T}_S^{ab}$ in Eq.\ \eqref{MC 2 forms special galileon} is symmetric), we are led to consider the most general invariant two-derivative terms built out of the relevant covariant objects, namely the components of the Maurer--Cartan 2-form, the covariant derivative $\nabla_a\sigma^{bc}$, as well as the second covariant derivative of the Goldstone $\xi^a$. We should stress that, in restricting our attention to those terms that are invariant under all the symmetries, despite the fact that the treating $e^a$ as non-dynamical breaks some of them, we are simply following the same strategy that yields ghost-free kinetic terms for massive spin-1 and spin-2 fields---{\it cf.}~comment in footnote \ref{footnote kinetic terms}.

This strategy yields in principle a slew of possible combinations, and we are faced with the problem of either (1) finding a specific choice (or set of choices) of operator coefficients such that the field $\sigma^{ab}$ can be completely removed from the kinetic terms, as we did for the potential terms; or (2) in the absence of such a choice, to determine if an action exists for both the spin-2 field $\omega_Q^a$ and the Goldstone $\sigma^{ab}$ that is free of pathologies. Although a full analysis is beyond the scope of this paper, a positive initial observation is that one can achieve the decoupling of $\sigma^{ab}$ at linear level when expanding the fields in perturbations about flat space.

To prove this, it is sufficient to note that the Goldstone covariant derivatives already furnish, at linear order about flat space, all the possible combinations of kinetic operators for $\sigma^{ab}$ and the metric fluctuation $\gamma^a$, the latter being defined by $q^a=\delta^a+\frac{1}{2}\,\gamma^a+O(\gamma^2)$. Notice that, by expanding the original definition $q^a=E^a+\Omega_Q^a$, we have that $\gamma^a=2\left(\sigma^a_{\phantom{a}b}\delta^b+(\omega^a_Q)^{(1)}\right)$, where $(\omega^a_Q)^{(1)}$ denotes the first-order piece in $\omega^a_Q$. In terms of $\gamma^a$ we have
\beq
\nabla_a\xi^b=(E^{-1})_b^{\phantom{b}\mu}(\Omega^a_Q)_{\mu}= \frac{1}{2}\,\gamma^a_{\phantom{a}b}+\cdots\,,
\eeq
where $\gamma^a_{\phantom{a}b}\equiv (\gamma^a)_{\mu}\delta^{\mu}_b$, and the ellipses stand for non-linear terms (which involve both $\gamma^a$ and $\sigma^{ab}$). Thus, the Goldstone covariant derivatives
\beq
\nabla_a\nabla_b\xi_c=\frac{1}{2}\,\partial_a\gamma_{bc}+\cdots\,,\qquad \nabla_a\sigma^{bc}=\partial_a\sigma_{bc}+\cdots\,,
\eeq
are sufficient to construct {\it any} set of kinetic operators in the action at quadratic order. In particular, we are free to engineer an action in which the Goldstone $\sigma^{ab}$ disappears from the kinetic terms at the quadratic level (while keeping the standard Fierz--Pauli kinetic terms for $\gamma_{ab}$).

Although this simple observation is encouraging, we have no reason to expect this decoupling to persist beyond linear order. Therefore, a more relevant question is whether the presence of $\sigma^{ab}$ necessarily leads to pathologies at the non-linear level. Notice that, if we were to choose to keep $\sigma^{ab}$ as a dynamical field, writing a healthy quadratic kinetic term for it is indeed possible, {\it i.e.}
\beq
S^{(2)}_{{\rm kin},\sigma}\propto \int d^Dx\bigg(-\frac{1}{2}\,\partial_a\sigma^{bc}\partial^a\sigma_{bc}+\partial_a\sigma^{bc}\partial_b\sigma^a_{\phantom{a}c}\bigg)\,.
\eeq
We have checked that this is the unique two-derivative quadratic action without ghosts for a traceless symmetric field. Unsurprisingly, this is nothing but the usual kinetic term for a spin-2 field, except that the trace of $\sigma$ is zero from the outset. This latter fact however does not affect the possible choices of kinetic operators that are free of ghosts.

To summarize, we have investigated how theories of massive gravity can be obtained from the gauging of the special galileon. The additional symmetry of the special galileon group, which by definition is spontaneously broken, gives rise to an extra Goldstone boson $\sigma^{ab}$. Interestingly, by restricting the spectrum of the gauge theory to contain only a massive spin-2 particle at zeroth order in derivatives, we showed that the symmetries single out a unique action (at least among the ghost-free class of potentials). We have also shown that the decoupling of the additional Goldstone can be achieved at the two-derivative level when expanding the action to quadratic order in perturbations about flat space. It remains to be seen whether such decoupling can be attained at the fully non-linear level, as was possible for the potentials, or at the very least whether it is possible to build an interacting action that is free of pathologies.

%%%%%%%%%%%%%%%%%%%%%%%%%%%%%%%%%%%%%%%%%%%
%%%%%%%%%%%%%%%%%%%%%%%%%%%%%%%%%%%%%%%%%%%

\section{Discussion} \label{sec:discussion}

In this paper we have discussed the gauging of non-linearly realized symmetries as a method to systematically construct spontaneously broken gauge theories. More specifically, we have addressed the question of how to derive a gauge theory in the Higgs phase from the knowledge of the Goldstone theory that it corresponds to in the decoupling limit. We have put forth the coset construction, along with its extension to include gauge symmetries, as a very general and systematic method to tackle the problem. Focusing on the particular case of the galileon shift symmetry, we have argued that its gauging may be used to investigate infrared completions of galileon theory, the goal being to better understand how massive gravity can be formulated when viewed as a gauge theory for the spontaneously broken diffeomorphism invariance.

%Our results followed from a number of assumptions made on physical grounds, and it may be interesting to modify or relax some of them. For instance, the introduction of gauge fields depended on which symmetries we wished to gauge, and we remarked that the choices we made were not unique. Particularly intriguing would be to consider the gauging of the extended shift symmetry of the special galileon theory. In this setting the additional Goldstone associated to the breaking of this symmetry would not by itself pose a problem, being now a pure gauge degree of freedom that would disappear in unitary gauge. However, the issue would then be to understand the possible interactions induced by the extra gauge field, a non-trivial task given that this field---a tensor-valued 1-form---would a priori contain a spin-3 degree of freedom upon expanding in perturbations around flat space.

Our results followed from a number of assumptions made on physical grounds, and it may be interesting to modify or relax some of them. First there is the question of which symmetries one wishes to gauge, and we remarked that the choices we made were not unique. This requires some physical input since formal consistency only demands that gauge symmetries make a subgroup. For the standard galileon we were naturally led to consider the gauging of the whole group: we insisted on gauging the galileon symmetries because of our interest in broken gauge theories, and we insisted on gauging Lorentz because of our wish to have healthy spin-2 kinetic terms. For the special galileon there is, in addition, the option of gauging the extended shift symmetry generated by $S_{ab}$. Although we chose not to do so, it would certainly be intriguing to consider this possibility. In this setting the additional Goldstone $\sigma^{ab}$ associated to the breaking of the extra symmetry would not by itself pose a problem, being now a pure gauge degree of freedom that would disappear in unitary gauge. However, the issue would then be to understand the possible interactions induced by the extra gauge field, a non-trivial task given that this field---a tensor-valued 1-form---would a priori contain a spin-3 degree of freedom upon expanding in perturbations around flat space.

A second assumption needed in the implementation of the coset construction concerns the inverse Higgs constraints. Typically, as we have explained, the fields one removes via such constraints are not restricted to be gapless by the symmetries. They are therefore massive in the absence of fine tuning, and one is allowed to integrate them out in order to focus on the gapless modes. This was the situation in our treatment of the standard galileon, where the Goldstone associated to the broken galileon shift could be removed---at low energies---without loss of generality. On the other hand, in the case of the special galileon, our insisting on having a potential with the dRGT structure led us to a theory where the Goldstone $\sigma^{ab}$ was {\it gapless}, despite the fact that the symmetries allowed for mass terms and the option of removing $\sigma^{ab}$ via an inverse Higgs constraint. Although it is likely that our choice is the only one that leads to a ghost-free potential for the graviton, we cannot discard the logical possibility that other potentials may exist for a {\it gapped} Goldstone $\sigma^{ab}$ which, after integrating out the latter, could produce a ghost-free mass term for the graviton.

Lastly, the derivation of our models relied on additional physical assumptions that are independent of the symmetry considerations dictated by the coset construction. One such assumption was the torsion-free condition that we chose to impose in both the standard and special galileon cases. Even though it was natural for us to avoid a dynamical torsion based on the spectrum we were interested in, it could be intriguing to eventually consider a Palatini-type formulation of our construction with the spin connection left unconstrained. Another assumption, based again on our wish to build a theory of massive gravity, was to set the vielbein associated to diffeomorphisms as a non-dynamical reference field. It would be of course natural to remove this constraint should one be interested in models of bi-gravity, but this would require some extra symmetries in order to produce the ingredients needed for building an additional spin-2 kinetic term.

Given the generality of the method, it is clear that this work can be extended in several directions, some of which we hope to address in future investigations. For instance, our analysis may be extended to the case of multi-galileons \cite{Padilla:2010de,Padilla:2010ir,Hinterbichler:2010xn} with the goal of exploring theories of multi-gravity \cite{Hinterbichler:2012cn,Hassan:2012wt,Hassan:2018mcw}. The extension to multiple scalars would also allow for the inclusion of additional internal symmetries \cite{Andrews:2010km,Allys:2016hfl}, thus serving as a starting point to investigate theories of massive spin-2 fields with extra symmetries of this type. Going beyond the standard galileon, it would be interesting to consider other related theories such as the conformal galileon and the DBI galileon \cite{deRham:2010eu}. Also intriguing would be the application of our techniques to Goldstone theories for which little intuition is available regarding the nature of the corresponding gauge theories, such as the $p$-form and tensor galileons \cite{Deffayet:2010zh,Deffayet:2016von,Deffayet:2017eqq,Chatzistavrakidis:2016dnj}. Lastly, relativistic versions of the galileon group and its symmetry breaking pattern---for example $ISO(3,1)\times ISO(3,1)\to ISO(3,1)$---could be helpful to address the problem of deriving ghost-free kinetic terms for multiple spin-2 fields in our formalism.

%%%%%%%%%%%%%%%%%%%%%%%%%%%%%%%%%%%%%%%%%%%
%%%%%%%%%%%%%%%%%%%%%%%%%%%%%%%%%%%%%%%%%%%

\begin{acknowledgments}

We would like to thank Luca Delacr\'etaz for important collaboration during the early stages of this project, and Alberto Nicolis for continuous encouragement during the long gestation period of this work. We are also grateful to Tomas Brauner, Mariana Carillo Gonz\'alez, Kurt Hinterbichler, Austin Joyce and Rachel A.\ Rosen for useful comments and discussions. SGS is supported by the European Research Council under the European Community's Seventh Framework Programme (FP7/2007-2013 Grant Agreement no.\ 307934, NIRG project); he also wishes to thank Carnegie Mellon University for generous hospitality. JK was supported by the Kwanjeong Educational Foundation.

\end{acknowledgments}

\appendix

\section{Ghost-free potentials in the gauged special galileon} \label{sec:app ghost free pot}

In this appendix we give a proof that the potential term
\begin{equation} \label{eq:appA_1}
S=\int\sum_{n=0}^D\frac{b_n}{n!(D-n)!}\,{\epsilon}_{a_1a_2\cdots a_D}\Omega_Q^{a_1}\wedge\cdots\wedge\Omega_Q^{a_n}\wedge E^{a_{n+1}}\wedge\cdots\wedge E^{a_D}\,,
\end{equation}
with
\beq\bal
E^a&=(\cosh\sigma)^a_{\phantom{a}b}e^b+(\sinh\sigma)^a_{\phantom{a}b}H^b\,,\\
\Omega_Q^a&=(\sinh\sigma)^a_{\phantom{a}b}e^b+(\cosh\sigma)^a_{\phantom{a}b}H^b\,,\\
\eal\eeq
is in fact independent of the Goldstone field $\sigma$ for the specific choice of coefficients
\beq \label{eq:appA_coeffs}
b_n=\begin{cases}
\beta_1 & \mbox{if $n$ is even}\,,\\
\beta_2 & \mbox{if $n$ is odd}\,,
\end{cases}
\eeq
and only for this choice. Although it is possible to show this by a direct calculation, we will employ an alternative method that will also turn out to be useful in Appendix~\ref{sec:app special gal}.

For simplicity, we will focus on the even terms in (\ref{eq:appA_1}), namely
\beq
S_1=\int\sum_{n\,{\mathrm{even}}}\frac{1}{n!(D-n)!}\,{\epsilon}_{a_1a_2\cdots a_D}\Omega_Q^{a_1}\wedge\cdots\wedge\Omega_Q^{a_n}\wedge E^{a_{n+1}}\wedge\cdots\wedge E^{a_D}\,,
\eeq
as the proof for the odd terms is completely analogous. It suffices to prove that $S_1$ remains unchanged upon making an infinitesimal variation of $\sigma$. Using that $\delta_{\sigma}E^a=\delta\sigma^a_{\phantom{a}b}\Omega_Q^b$ and $\delta_{\sigma}\Omega_Q^a=\delta\sigma^a_{\phantom{a}b}E^b$, we have
\beq\bal \label{eq:appA_2}
\delta_{\sigma} S_1&=\int\sum_{n\,{\mathrm{even}}}\frac{1}{n!(D-n)!}\,{\epsilon}_{a_1a_2\cdots a_D}\bigg[n\,\delta\sigma^{a_1}_{\phantom{a_1}b}E^b\wedge\Omega_Q^{a_2}\wedge\cdots\wedge\Omega_Q^{a_n}\wedge\omega_P^{a_{n+1}}\wedge\cdots\wedge E^{a_D}\\
&\quad+(D-n)\,\delta\sigma^{a_1}_{\phantom{a_1}b}\Omega_Q^b\wedge\Omega_Q^{a_2}\wedge\cdots\wedge\Omega_Q^{a_{n+1}}\wedge E^{a_{n+2}}\wedge\cdots\wedge E^{a_D}\bigg]\\
&=\int\sum_{n\,{\mathrm{even}}}{\epsilon}_{a_1a_2\cdots a_D}\bigg[\frac{1}{n!(D-n-1)!}\,\delta\sigma^{a_1}_{\phantom{a_1}b}\Omega_Q^b\wedge\Omega_Q^{a_2}\wedge\cdots\wedge\Omega_Q^{a_{n+1}}\wedge E^{a_{n+2}}\wedge\cdots\wedge E^{a_D}\\
&\quad+\frac{1}{(n+1)!(D-n-2)!}\,\delta\sigma^{a_1}_{\phantom{a_1}b}\omega_P^b\wedge\Omega_Q^{a_2}\wedge\cdots\wedge\Omega_Q^{a_{n+2}}\wedge E^{a_{n+3}}\wedge\cdots\wedge E^{a_D}\bigg]\\
&=\int\sum_{n\,{\mathrm{even}}}\frac{{\epsilon}_{a_1a_2\cdots a_D}}{(n+1)!(D-n-1)!}\bigg[(n+1)\,\delta\sigma^{a_1}_{\phantom{a_1}b}\Omega_Q^b\wedge\Omega_Q^{a_2}\wedge\cdots\wedge\Omega_Q^{a_{n+1}}\wedge E^{a_{n+2}}\wedge\cdots\wedge E^{a_D}\\
&\quad+(D-n-1)\,\delta\sigma^{a_1}_{\phantom{a_1}b}\omega_P^b\wedge\Omega_Q^{a_2}\wedge\cdots\wedge\Omega_Q^{a_{n+2}}\wedge E^{a_{n+3}}\wedge\cdots\wedge E^{a_D}\bigg]\,.\\
\eal\eeq
Next we show that the term in square brackets in the last line vanishes for all $n$. Consider the following expression:
\begin{equation}
\begin{split}
&{\epsilon}_{a_1a_2\cdots a_D}\,\delta\sigma^{a_1}_{\phantom{a_1}b}\Omega_Q^b\wedge\Omega_Q^{a_2} \wedge \cdots \wedge \Omega_Q^{a_{n}}\wedge E^{a_{n+1}}\wedge\cdots\wedge E^{a_D}\\
&\quad={\epsilon}_{a_1a_2\cdots a_D}\delta^{b_1}_b\,\delta\sigma^{a_1}_{\phantom{a_1}b_1}\Omega_Q^b\wedge\Omega_Q^{a_2}\wedge\cdots\wedge\Omega_Q^{a_{n}}\wedge E^{a_{n+1}}\wedge\cdots\wedge E^{a_D}\\
&\quad=-\frac{1}{(D-1)!}\,{\epsilon}_{a_1a_2\cdots a_D}{\epsilon}^{b_1b_2\cdots b_D}{\epsilon}_{bb_2\cdots b_D}\,\delta\sigma^{a_1}_{\phantom{a_1}b_1}\Omega_Q^b\wedge\Omega_Q^{a_2}\wedge\cdots\wedge\Omega_Q^{a_{n}}\wedge E^{a_{n+1}}\wedge\cdots\wedge E^{a_D}\,.\\
\end{split}
\end{equation}
Using the identity
\beq\bal
{\epsilon}_{a_1a_2\cdots a_D}{\epsilon}^{b_1b_2\cdots b_D}{\epsilon}_{bb_2\cdots b_D}&=-\delta^{b_1\cdots b_D}_{a_1\cdots a_D}{\epsilon}_{bb_2\cdots b_D}\\
&=-\bigg[\delta^{b_1}_{a_1}\delta^{b_2\cdots b_D}_{a_2\cdots a_D}-\sum_{k=2}^{D}\delta^{b_1}_{a_k}\delta^{b_2\cdots b_D}_{a_2\cdots a_{k-1}a_1 a_{k+1}\cdots a_D}\bigg]{\epsilon}_{bb_2\cdots b_D}\\
&=-(D-1)!\bigg[\delta^{b_1}_{a_1}{\epsilon}_{ba_2\cdots a_D}-\sum_{k=2}^{D}\delta^{b_1}_{a_k}{\epsilon}_{ba_2\cdots a_{k-1}a_1 a_{k+1}\cdots a_D}\bigg]\,,
\eal\eeq
and the fact that $\delta\sigma$ is traceless, we infer that
\begin{equation}
\begin{split}
&{\epsilon}_{a_1a_2\cdots a_D}\,\delta\sigma^{a_1}_{\phantom{a_1}b}\Omega_Q^b\wedge\Omega_Q^{a_2}\wedge\cdots\wedge\Omega_Q^{a_{n}}\wedge E^{a_{n+1}}\wedge\cdots\wedge E^{a_D}\\
&\quad=-\sum_{k=2}^{D}{\epsilon}_{ba_2\cdots a_{k-1}a_1 a_{k+1}\cdots a_D}\,\delta\sigma^{a_1}_{\phantom{a_1}a_k}\Omega_Q^b\wedge\Omega_Q^{a_2}\wedge\cdots\wedge\Omega_Q^{a_{n}}\wedge E^{a_{n+1}}\wedge\cdots\wedge E^{a_D}\\
&\quad=-(n-1)\,{\epsilon}_{a_1a_2\cdots a_D}\,\delta\sigma^{a_1}_{\phantom{a_1}b}\Omega_Q^b\wedge\Omega_Q^{a_2}\wedge\cdots\wedge\Omega_Q^{a_{n}}\wedge E^{a_{n+1}}\wedge\cdots\wedge E^{a_D}\\
&\quad-(D-n)\,{\epsilon}_{a_1a_2\cdots a_D}\,\delta\sigma^{a_1}_{\phantom{a_1}b} E^b\wedge\Omega_Q^{a_2}\wedge\cdots\wedge\Omega_Q^{a_{n+1}}\wedge E^{a_{n+2}}\wedge\cdots\wedge E^{a_D}\,,
\end{split}
\end{equation}
that is
\begin{equation}
\begin{split}
&n\,{\epsilon}_{a_1a_2\cdots a_D}\,\delta\sigma^{a_1}_{\phantom{a_1}b}\Omega_Q^b\wedge\Omega_Q^{a_2}\wedge\cdots\wedge\Omega_Q^{a_{n}}\wedge E^{a_{n+1}}\wedge\cdots\wedge E^{a_D}\\
&\quad=-(D-n)\,{\epsilon}_{a_1a_2\cdots a_D}\,\delta\sigma^{a_1}_{\phantom{a_1}b} E^b\wedge\Omega_Q^{a_2}\wedge\cdots\wedge\Omega_Q^{a_{n+1}}\wedge E^{a_{n+2}}\wedge\cdots\wedge E^{a_D}\,.
\end{split}
\end{equation}
Using this last result with $n \to n+1$ in the last line of eq.\ \eqref{eq:appA_2} establishes that $\delta_{\sigma} S_1=0$. Repeating the same steps for the odd terms in the action we reckon that $\delta_{\sigma} S=0$, so that the full action $S$ is indeed independent of $\sigma$.

The fact that the coefficients \eqref{eq:appA_coeffs} are the {\it unique} possible choice follows simply by construction. The variation of the $n=0$ term in $S$ yields a term with one factor of $E$, which can only be canceled by the variation of the $n=2$ term. But this term will also include an expression with three factors of $E$, which again must be matched with one coming from the $n=4$ term. This construction thus produces a unique set of coefficients $b_n$ for all even $n$. The same argument of course applies to the odd-$n$ coefficients as well.

%%%%%%%%%%%%%%%%%%%%%%%%%%%%%%%%%%%%%%%%%%%
%%%%%%%%%%%%%%%%%%%%%%%%%%%%%%%%%%%%%%%%%%%

\section{Coset construction of the special galileon algebra} \label{sec:app special gal}

The derivation of the special galileon from a coset construction has been first discussed in~\cite{Bogers:2018zeg}. Given that several steps of the calculation are quite non-trivial and are not shown in~\cite{Bogers:2018zeg}, we deem it useful to repeat it here in some detail. Along the way, we will also generalize this construction to arbitrary dimensions.

Let us start by summarizing all the non-vanishing commutators of the algebra we are interested in:
\beq\bal
\left[P_a,Q_b\right]&=\eta_{ab}C\,,\\
[J_{ab},Q_c]&=\eta_{bc}Q_a-\eta_{ac}Q_b\,,\\
[J_{ab},P_c]&= \eta_{bc} P_a - \eta_{ac} P_b\,,\\
[J_{ab},J_{cd}]&=\eta_{bc} J_{ad} - \eta_{ac} J_{bd} - \eta_{bd} J_{ac} + \eta_{ad} J_{bc}\,,\\
\left[P_a,S_{bc}\right]&=\eta_{ab}Q_c+\eta_{ac}Q_b-\frac{2}{D}\,\eta_{bc}Q_a\,,\\
[Q_a,S_{bc}]&=\alpha^2\left(\eta_{ab}P_c+\eta_{ac}P_b-\frac{2}{D}\,\eta_{bc}P_a\right)\,,\\
[S_{ab},S_{cd}]&=\alpha^2\left(\eta_{bc} J_{ad}+\eta_{ac} J_{bd} + \eta_{bd} J_{ac} + \eta_{ad} J_{bc}\right)\,,\\
[J_{ab},S_{cd}]&=\eta_{bc}S_{ad}-\eta_{ac}S_{bd}+\eta_{bd}S_{ac}-\eta_{ad}S_{bc}\,,
\eal\eeq
(for the sake of clarity we will keep $\alpha$ explicit now). The Goldstone fields are then introduced as the parameters that enter the coset representative
\beq \label{coset App B}
\Omega=e^{x^aP_a}e^{\pi C}e^{\xi^bQ_b}e^{\frac{1}{2}\sigma^{cd}S_{cd}}\,.
\eeq

\subsection{Maurer--Cartan form}

The Maurer--Cartan form is expanded as
\beq \label{eq:mc form global special gal}
\Omega^{-1}\d\Omega=E^aP_a+\frac{1}{2}\,\Omega^{ab}J_{ab}+\Omega_Q^aQ_a+\Omega_CC+\frac{1}{2}\,\Omega_S^{ab}S_{ab}\,,
\eeq
where
\beq\bal \label{eq:covariant forms special gal}
E^a&=(\cosh\alpha\sigma)^{a}_{\phantom{a}b}\d x^b+\alpha(\sinh\alpha\sigma)^{a}_{\phantom{a}b}\d\xi^b\,,\\
\Omega^{ab}&=\alpha^2(\Lambda^{-1})^{ab}_{cd}\left(\cosh\alpha\Lambda-\mathbf{1}\right)^{cd}_{ef}\d\sigma^{ef} \,,\\
\Omega_Q^a&=\frac{1}{\alpha}(\sinh\alpha\sigma)^{a}_{\phantom{a}b}\d x^b+(\cosh\alpha\sigma)^{a}_{\phantom{a}b}\d\xi^b\,,\\
\Omega_C&=\d\pi+\xi_a\d x^a\,,\\
\Omega_S^{ab}&=\frac{1}{\alpha}(\Lambda^{-1})^{ab}_{cd}\left(\sinh\alpha\Lambda\right)^{cd}_{ef}\d\sigma^{ef} \,,
\eal\eeq
with the definition $\Lambda^{ab}_{cd}\equiv \delta^a_c\sigma^b_{\phantom{b}d}-\delta^b_d\sigma^a_{\phantom{a}c}$. While the computation of $E^a$, $\Omega_Q^a$ and $\Omega_C$ is straightforward, obtaining the closed form expression for $\Omega^{ab}$ and $\Omega_S^{ab}$ shown above via a direct calculation is quite non-trivial. For this reason, we derived the explicit results in \eqref{eq:mc form global special gal} using a series of  tricks, which we will now explain in detail. 

We will start by replacing $\sigma\to\lambda\sigma$ in Eq.\ \eqref{coset App B}, and noting from Eq.\ \eqref{eq:mc form global special gal} that
\beq
\frac{\Omega_{S,\lambda}^{ab}}{2}\,S_{ab}+\frac{\Omega_{\lambda}^{ab}}{2}\,J_{ab}=e^{-\frac{\lambda}{2}\sigma\cdot S}\d e^{\frac{\lambda}{2}\sigma\cdot S}\,.
\eeq
The trick is now to differentiate with respect to $\lambda$ in order to get differential equations for $\Omega_{S,\lambda}^{ab}$ and $\Omega_{\lambda}^{ab}$. Using the algebra commutators we find
\beq\bal \label{eq:diff eqs special gal}
\frac{d\Omega_{S,\lambda}^{ab}}{d\lambda}&=-2\sigma^{(a}_{\phantom{(a}c}\Omega_{\lambda}^{|c|b)}+\d\sigma^{ab}\,,\\
\frac{d\Omega_{\lambda}^{ab}}{d\lambda}&=-2\alpha^2\sigma^{[a}_{\phantom{[a}c}\Omega_{S,\lambda}^{|c|b]}\,,
\eal\eeq
and the ``initial conditions'' are $\Omega_{S,\lambda=0}^{ab}=0$ and $\Omega_{\lambda=0}^{ab}=0$.\footnote{In the presence of gauge fields ({\it cf.}\ Sec.\ \ref{sec:special galileon}), the differential equations \eqref{eq:diff eqs special gal} would remain unchanged while the initial conditions would be modified.} Ignoring the $\d\sigma^{ab}$ term in \eqref{eq:diff eqs special gal} we have a set of homogeneous equations with general solution
\beq\bal
\Omega_{S,\lambda}^{ab}&=A_1^{cd}\left((\cosh\lambda\alpha\sigma)_c^{\phantom{c}a}(\cosh\lambda\alpha\sigma)_d^{\phantom{d}b}-(\sinh\lambda\alpha\sigma)_c^{\phantom{c}a}(\sinh\lambda\alpha\sigma)_d^{\phantom{d}b}\right)\\
&\quad+A_2^{cd}\left((\cosh\lambda\alpha\sigma)_c^{\phantom{c}a}(\sinh\lambda\alpha\sigma)_d^{\phantom{d}b}-(\sinh\lambda\alpha\sigma)_c^{\phantom{c}a}(\cosh\lambda\alpha\sigma)_d^{\phantom{d}b}\right)\,,\\
\Omega_{\lambda}^{ab}&=\alpha A_2^{cd}\left((\cosh\lambda\alpha\sigma)_c^{\phantom{c}a}(\cosh\lambda\alpha\sigma)_d^{\phantom{d}b}-(\sinh\lambda\alpha\sigma)_c^{\phantom{c}a}(\sinh\lambda\alpha\sigma)_d^{\phantom{d}b}\right)\\
&\quad\alpha A_1^{cd}\left((\cosh\lambda\alpha\sigma)_c^{\phantom{c}a}(\sinh\lambda\alpha\sigma)_d^{\phantom{d}b}-(\sinh\lambda\alpha\sigma)_c^{\phantom{c}a}(\cosh\lambda\alpha\sigma)_d^{\phantom{d}b}\right)\,,\\
\eal\eeq
where $A_1^{ab}$ is symmetric and traceless and $A_2^{ab}$ is antisymmetric. On the other hand, a particular solution of \eqref{eq:diff eqs special gal} is provided by
\beq
\Omega_{S,\lambda}^{ab}=0\,,\qquad \Omega_{\lambda}^{ab}=B^{ab}\,,
\eeq
where $B^{ab}$ (a 1-form) is independent of $\lambda$ and satisfies, in matrix notation,
\beq\label{eq:Beq special gal}
[\sigma,B]=\d\sigma\,.
\eeq
This equation cannot in general be solved explicitly for $B$. Nevertheless, the final result for the Maurer--Cartan form components can be written in such a way that $B$ disappears, as we will now show.

First, from the initial conditions we find $A_1^{ab}=0$ and $A_2^{ab}=-\frac{1}{\alpha}\,B^{ab}$, so that setting at last $\lambda=1$ we have the desired solutions
\begin{align}
\Omega_{S}^{ab}&=-\frac{1}{\alpha}\,B^{cd}\left((\cosh\alpha\sigma)_c^{\phantom{c}a}(\sinh\alpha\sigma)_d^{\phantom{d}b}-(\sinh\alpha\sigma)_c^{\phantom{c}a}(\cosh\alpha\sigma)_d^{\phantom{d}b}\right)\,, \label{Omega S app B} \\
\Omega^{ab}&=B^{ab}-B^{cd}\left((\cosh\alpha\sigma)_c^{\phantom{c}a}(\cosh\alpha\sigma)_d^{\phantom{d}b}-(\sinh\alpha\sigma)_c^{\phantom{c}a}(\sinh\alpha\sigma)_d^{\phantom{d}b}\right)\, .
\end{align}
Switching again to matrix notation, we can expand the RHS of \eqref{Omega S app B} in powers of $\sigma$ and recast it as
\beq \label{Omega S app B 2}
\Omega_S^{ab}=\sum_{m,n=0}^{\infty}\frac{\alpha^{2(m+n)}}{(2m+1)!(2n)!}\left(\sigma^{2m+1}B\sigma^{2n}-\sigma^{2n}B\sigma^{2m+1}\right)^{ab}\,.
\eeq
Next, from Eq.\ \eqref{eq:Beq special gal} we infer that
\beq
[\sigma^k,B]=\sum_{j=0}^{k-1}\sigma^j\d\sigma \sigma^{k-1-j}\,,
\eeq
which allows us to rewrite \eqref{Omega S app B 2} as
\beq\bal \label{eq:omegaS result special gal}
\Omega_S^{ab}&=\sum_{m,n=0}^{\infty}\frac{\alpha^{2(m+n)}}{(2m+1)!(2n)!}\left([\sigma^{2m+1},B]\sigma^{2n}-[\sigma^{2n},B]\sigma^{2m+1}\right)^{ab}\\
&=\sum_{k=0}^{\infty}\frac{\alpha^{2k}}{(2k+1)!}\sum_{l=0}^{2k}(-1)^l\left(\begin{matrix}2k\\ l\end{matrix}\right) \left(\sigma^l\d\sigma \sigma^{2k-l}\right)^{ab}\, .
\eal\eeq
This result is manifestly independent of the unknown matrix $B$, as claimed above. 

In order to find a closed form expression for $\Omega_S^{ab}$, we can use the binomial theorem to find
\beq
\sum_{l=0}^{2k}(-1)^l\left(\begin{matrix}2k\\ l\end{matrix}\right) \left(\sigma^l\d\sigma \sigma^{2k-l}\right)^{ab}=(\Lambda^{2k})^{ab}_{cd}\,\d\sigma^{cd}\,,\qquad \Lambda^{ab}_{cd}\equiv \delta^a_c\sigma^b_{\phantom{b}d}-\delta^b_d\sigma^a_{\phantom{a}c}\,.
\eeq
Substituting this expression in Eq.\ \eqref{eq:omegaS result special gal} gives the result quoted in \eqref{eq:covariant forms special gal} after some manipulations. Lastly, having found $\Omega_S^{ab}$, we can determine $\Omega^{ab}$ from the differential equation \eqref{eq:diff eqs special gal} by direct integration. This yields the result shown in \eqref{eq:covariant forms special gal}.

\subsection{Inverse Higgs constraints}

For completeness we also include here the solution of the inverse Higgs constraints, even though they may be easily obtained from our results of Sec.\ \ref{sec:special galileon} for the gauged special galileon. The standard first constraint inferred from the algebra commutator $\left[P_a,Q_b\right]=\eta_{ab}C$ implies
\beq
\nabla_a\pi=(E^{-1})_a^{\phantom{a}\mu}(\partial_{\mu}\pi+\xi_{\mu})=0\qquad \Rightarrow\qquad \xi_a=-\partial_a\pi\,,
\eeq
remembering that $\partial_a\equiv (E^{-1})_a^{\phantom{a}\mu}\partial_{\mu}$. The second constraint arises from the commutator $\left[P_a,S_{bc}\right]=\eta_{ab}Q_c+\eta_{ac}Q_b-\frac{2}{D}\,\eta_{bc}Q_a$, which gives the following relation for the covariant derivative of the Goldstone $\xi_a$:
\beq \label{eq:second ihc appendix}
\nabla_a\xi_b+\nabla_b\xi_a-\frac{2}{D}\,\eta_{ab}[\nabla\xi]=0\,,
\eeq
where $[\ldots]$ denotes the trace in matrix notation. From \eqref{eq:covariant forms special gal} we have
\beq
\nabla\xi=(\cosh\sigma+K\cdot\sinh\sigma)^{-1}\cdot (\sinh\sigma+K\cdot\cosh\sigma)\,,
\eeq
where $K_a^{\phantom{a}b}\equiv \partial_a\xi^b$. The last expression may be simplified by noting that we must have $[\sigma,K]=0$ by consistency, since the inverse Higgs constraint, Eq.\ \eqref{eq:second ihc appendix}, ultimately implies a relation between $\sigma$ and the matrix $K$. It then follows that
\beq \label{eq:second ihc appendix 2}
\nabla\xi=\tanh(\sigma+\tanh^{-1}K)\,.
\eeq
We now apply the first constraint to infer that $K_{ab}=-\partial_a\partial_b\pi$, and thus the matrix $\nabla\xi$ is actually symmetric. Eq.\ \eqref{eq:second ihc appendix} then states that $\nabla\xi=\frac{\mathbf{1}}{D}\,[\nabla\xi]$, which can be combined with \eqref{eq:second ihc appendix 2} to yield the desired solution for the Goldstone $\sigma^{ab}$ as a function of $\pi$,
\beq
\sigma^{ab}=-(\tanh^{-1}K)^{ab}+\frac{1}{D}\,\eta^{ab}[\tanh^{-1}K]\,.
\eeq

\subsection{Wess--Zumino terms}

As already mentioned in Sec.\ \ref{sec:coset}, the covariance of the coset vielbein and the Goldstone covariant derivatives ensures that $H$-invariant contractions will be invariant under the whole group $G$. In addition, there exists also the possibility to include ``Wess--Zumino (WZ) terms'' (also called Chern--Simons terms in other contexts) in the effective action. WZ terms are by definition invariant under the symmetries only modulo a total derivative, and cannot be written as an exactly invariant expression upon integrating by parts. In mathematical language, they are in one-to-one correspondence with the elements of the relative Lie algebra cohomology of $G/H$ in the space of $G$-valued $(D+1)$-forms (with $D$ the spacetime dimension). We will now show how to carry out such construction in the case of the special galileon. This turns out to be a particularly interesting application, because the leading-order Lagangian for the special galileon is precisely a WZ term.

The calculation starts by determining the elements of the relative Lie algebra cohomology. This can be done solely from the knowledge of the algebra, from which the Cartan structure equation follows: $\d\Theta=-\frac{1}{2}\,\Theta^2$, where $\Theta$ is the Maurer--Cartan form. In components this gives, for the special galileon,
\beq\bal \label{eq:cartan eqs special gal}
\d E^a&=\Omega^{\phantom{S}a}_{S\phantom{a}b}\wedge\Omega_Q^b\,,\\
\d\Omega_Q^a&=\Omega^{\phantom{S}a}_{S\phantom{a}b}\wedge E^b\,,\\
\d\Omega_C&=-E_a\wedge\Omega_Q^b\,,\\
\d\Omega_S^{ab}&=0\,,
\eal\eeq
and we are ignoring terms involving $\Omega^{ab}$ on the right-hand side, as such terms must necessarily cancel in any Lorentz invariant quantity. In the case of the {\it standard} galileon the WZ terms are obtained from the $(D+1)$-forms \cite{Goon:2012dy}
\beq
\gamma_n\equiv {\epsilon}_{a_1\cdots a_D}\Omega_C\wedge\Omega_Q^{a_1}\wedge\cdots\wedge\Omega_Q^{a_n}\wedge E^{a_{n+1}}\wedge\cdots\wedge E^{a_D}\,,
\eeq
which are $\d$-exact for any $n=0,\ldots,D$. This is not true anymore for the special galileon, because the presence of $\Omega_S$ on the right-hand side of \eqref{eq:cartan eqs special gal} implies that $\d\gamma_n\neq0$. There are however two particular linear combinations of the $\gamma_n$ forms that do close under $\d$, hence giving rise to two candidate WZ terms.

To this end we write the most general combination of the above forms as
\beq
\gamma=\sum_{n=0}^D\frac{a_n}{n!(D-n)!}\,{\epsilon}_{a_1a_2\cdots a_D}\Omega_C\wedge\Omega_Q^{a_1}\wedge\cdots\wedge\Omega_Q^{a_n}\wedge E^{a_{n+1}}\wedge\cdots\wedge E^{a_D}\,.
\eeq
The goal is now to determine the set of coefficients $a_n$ for which $\d\gamma=0$. Upon taking the exterior derivative the term proportional to $\d\Omega_C$ vanishes, so that after some manipulations we are left with
\beq\bal
\d\gamma&=-\sum_{n=0}^D\frac{a_n}{n!(D-n)!}\,{\epsilon}_{a_1a_2\cdots a_D}\Omega_C\wedge\bigg[n\,\Omega^{\phantom{S}a_1}_{S\phantom{a_1}b}\wedge E^b\wedge\Omega_Q^{a_2}\wedge\cdots\wedge\Omega_Q^{a_n}\wedge E^{a_{n+1}}\wedge\cdots\wedge E^{a_D}\\
&\quad+(D-n)\,\Omega^{\phantom{S}a_1}_{S\phantom{a_1}b}\wedge\Omega_Q^b\wedge\Omega_Q^{a_2}\wedge\cdots\wedge\Omega_Q^{a_{n+1}}\wedge E^{a_{n+2}}\wedge\cdots\wedge E^{a_D}\bigg]\,.\\
\eal\eeq
This result is formally identical to the expression one obtains upon varying the dRGT type potential, Eq.\ \eqref{eq:appA_1}, with respect to $\sigma$ (\textit{cf.}~\eqref{eq:appA_2}), and which we used as the basis of our proof in Appendix \ref{sec:app ghost free pot}. The only difference is that now $\Omega_S$ appears instead of $\sigma$, but because both $\Omega_S$ and $\sigma$ are symmetric and traceless, all the steps performed in Appendix \ref{sec:app ghost free pot} can be repeated verbatim here to conclude that $\d\gamma=0$ if, and only if, the coefficients $a_n$ satisfy
\beq
a_n=\begin{cases}
\alpha_1 & \mbox{if $n$ is odd}\,,\\
\alpha_2 & \mbox{if $n$ is even}\,.
\end{cases}
\eeq
There are therefore precisely two candidate WZ terms,
\beq\bal
\Gamma_1&=\sum_{n\,{\mathrm{odd}}}\frac{1}{n!(D-n)!}\,{\epsilon}_{a_1a_2\cdots a_D}\Omega_C\wedge\Omega_Q^{a_1}\wedge\cdots\wedge\Omega_Q^{a_n}\wedge E^{a_{n+1}}\wedge\cdots\wedge E^{a_D}\,,\\
\Gamma_2&=\sum_{n\,{\mathrm{even}}}\frac{1}{n!(D-n)!}\,{\epsilon}_{a_1a_2\cdots a_D}\Omega_C\wedge\Omega_Q^{a_1}\wedge\cdots\wedge\Omega_Q^{a_n}\wedge E^{a_{n+1}}\wedge\cdots\wedge E^{a_D}\,.
\eal\eeq
Finding the corresponding Lagrangian $D$-forms is simplified by the fact that the terms $\Gamma_{1,2}$ are actually independent of $\sigma^{ab}$, again from the results of Appendix \ref{sec:app ghost free pot}. From the expressions in \eqref{eq:covariant forms special gal} we thus have
\beq\bal
\Gamma_1&=\sum_{n\,{\mathrm{odd}}}\frac{1}{n!(D-n)!}\,{\epsilon}_{a_1a_2\cdots a_D}(\d\pi+\xi_a\d x^a)\wedge\d\xi^{a_1}\wedge\cdots\wedge\d\xi^{a_n}\wedge \d x^{a_{n+1}}\wedge\cdots\wedge\d x^{a_D}\,,\\
\Gamma_2&=\sum_{n\,{\mathrm{even}}}\frac{1}{n!(D-n)!}\,{\epsilon}_{a_1a_2\cdots a_D}(\d\pi+\xi_a\d x^a)\wedge\d\xi^{a_1}\wedge\cdots\wedge\d\xi^{a_n}\wedge \d x^{a_{n+1}}\wedge\cdots\wedge\d x^{a_D}\,.\\
\eal\eeq
Let us focus on the first term and write $\Gamma_1=\d \Lambda_1$. We then find that
\beq\bal
\Lambda_1&=\sum_{n\,{\mathrm{odd}}}\frac{1}{n!(D-n)!}\,{\epsilon}_{a_1a_2\cdots a_D}\bigg[\pi\d\xi^{a_1}\wedge\cdots\wedge\d\xi^{a_n}\wedge \d x^{a_{n+1}}\wedge\cdots\wedge\d x^{a_D}\\
&\quad-\frac{n}{2(D-n-1)}\,\xi^2\d\xi^{a_1}\wedge\cdots\wedge\d\xi^{a_{n-1}}\wedge \d x^{a_{n}}\wedge\cdots\wedge\d x^{a_D}\bigg]\,.
\eal\eeq
Finally, the Lagrangian is obtained from $\Lambda_1$ after pulling back to the $D$-dimensional spacetime manifold. At this stage we may apply the inverse Higgs constraint $\xi_a=-\partial_a\pi$ to arrive at
\beq
{\mathcal{L}}_1=\frac{1}{2}\sum_{n\,{\mathrm{odd}}}\frac{1}{n!}\,\pi\,\delta^{b_1\cdots b_n}_{a_1\cdots a_n}\partial_{b_1}\partial^{a_1}\pi\cdots\partial_{b_n}\partial^{a_n}\pi\,.
\eeq
We have thus reproduced the special galileon Lagrangian from the coset construction. The calculation of the Lagrangian that follows from $\Gamma_2$ is of course analogous and produces an independent term, which however has a tadpole and will usually be ignored in physical applications.

%%%%%%%%%%%%%%%%%%%%%%%%%%%%%%%%%%%%%%%%%%%
%%%%%%%%%%%%%%%%%%%%%%%%%%%%%%%%%%%%%%%%%%%

\bibliographystyle{apsrev4-1}
\bibliography{GaugedGalileonDraft}

\end{document}